\documentclass[journal]{new-aiaa_modified}
\usepackage[utf8]{inputenc}

\usepackage{graphicx}
\usepackage[version=4]{mhchem}
\usepackage{longtable,tabularx}
\usepackage{fancyhdr}
\usepackage{multicol}
\newenvironment{Figure}
{\par\medskip\noindent\minipage{\linewidth}}
{\endminipage\par\medskip}

\usepackage{caption}
\usepackage{subcaption}
\usepackage{xcolor}
\usepackage{txfonts}
 
\newcommand{\pde}[2]{\frac{\partial #1}{\partial #2}}

\title{Quasi-Two-Dimensional Simulation of a Rotating Detonation Engine Combustor and Injector}

\author{S. She-Ming Lau-Chapdelaine\footnote{Corresponding author, Assistant Professor, Department of Chemistry and Chemical Engineering, e-mail: shem.lau-chapdelaine@rmc.ca; \\ \hspace*{1cm}Work started as a post-doc at the National Research Council joint with the University of Ottawa and completed at the Royal Military College.}}
\affil{Royal Military College, Kingston, ON, K7K 7B4, Canada}
\author{Matei I. Radulescu\footnote{Professor, Department of Mechanical Engineering}}
\affil{University of Ottawa, Ottawa, ON, K1N 6N5, Canada}
\author{Zekai Hong\footnote{Research Officer, Aerospace Research Center, AIAA Senior Member}}
\affil{National Research Council, Ottawa, ON, K1A 0R6, Canada\vspace{8pt}\\{\normalfont\small \href{https://doi.org/10.2514/1.B39214}{https://doi.org/10.2514/1.B39214}}}

\begin{document}
\fancypagestyle{firststyle}
{
	\fancyhf{}
	\fancyhead[LO]{\textsc{Journal of Propulsion and Power}\\Vol. 40, No. 1, January--February 2024}
	\cfoot{42}
}	
\thispagestyle{firststyle}

\maketitle

\setcounter{page}{42}
\pagestyle{fancy}
\fancyhead{} 
\fancyfoot{}
\fancyhead[RO,LE]{\thepage}
\fancyhead[C]{\small\MakeUppercase{She-Ming Lau-Chapdelaine, Radulescu, and Hong}}

\begin{abstract}
	A numerical simulation of an annular rotating detonation engine with stoichiometric hydrogen-oxygen is performed. A generic, well posed, and easily implemented approach using a quasi-two-dimensional method to model the area variations through the rotating detonation engine's injector and combustor is presented. The detonation--injector interaction is studied for the case with a ratio of four between the combustor and injector's throat areas. A shock wave is formed in the divergent portion of the injector due the high back pressure created by the detonation in the combustor. A Favre-averaged steady-state analysis of stream lines and particle paths reveals that the shock causes an irrecoverable loss of stagnation pressure. Stagnation pressure gain in the combustor is insufficient to make up for the loss and the flow leaves the engine with lower stagnation pressure than in the plenum.
\end{abstract}

\footnotemark[0]\footnotetext[0]{
This is the authors' version of an article submitted to the Journal of Propulsion and Power on 2021-06-10, resubmitted on 2023-04-04, revised 2023-04-13, 2023-07-10, accepted 2023-07-16, and published online 2023-09-25. The published version is available as: S. S.-M. Lau-Chapdelaine, M. I. Radulescu, and Z. Hong. Quasi-Two-Dimensional Simulation of a Rotating Detonation Engine Combustor and Injector. Journal of Propulsion and Power. Vol. 40, No. 1, 2024, pp. 42--49. doi:\href{https://doi.org/10.2514/1.B39214}{https://doi.org/10.2514/1.B39214}. eISSN: 1533-3876
\\This work was presented at the AIAA Propulsion and Energy 2020 Forum, August 24-28, virtual event, AIAA 2020-3878, \href{https://doi.org/10.2514/6.2020-3878}{https://doi.org/10.2514/6.2020-3878}.
\\
Copyright © 2023 by His Majesty the King in Right of Canada, as represented by the Minister of Innovation, Science and Industry.}

\begin{multicols}{2}
\section{Introduction} 
\lettrine{T}{he} distinguishing feature of the annular rotating detonation engine (RDE) is its cylindrical annular combustor in which a detonation wave propagates azimuthally to the mean flow. The detonation is sustained by the axial injection of reactants from one end that produces a region of fresh reactants between wave fronts. Behind the detonation, pressure waves and combustion products can be driven into the injectors and plenum, temporarily blocking the entrance of reactants into the combustion chamber \cite{schwer2012feedback}. As pressure on the injection face drops, fresh gases refill the chamber until the detonation returns to consume them. High pressure and temperature products leave from the other end of the engine at high speeds, producing continuous thrust.

Many types of continuous detonation engines exist (\textit{e.g.} annular, hollow \cite{tangThreedimensionalNumericalInvestigations2015}, disk \cite{nakagamiExperimentalVisualizationStructure2017}, and shuttling \cite{yamaguchiSupersonicCombustionInduced2019a}) and promise large gains in efficiency \cite{heiser2002thermodynamic} premised on thermodynamic analyses of the combustor, but the injectors have seldom been considered in these predictions. The injectors are usually a configuration of holes, slots, or combinations of the two, with converging- or converging-diverging geometries. 
In numerical simulations, the injectors are often replaced by area-averaged boundary conditions \cite{zhdan1990calculation,yi2011propulsive}. These boundary conditions block backflow when pressure at the boundary is higher than the stagnation (total) pressure. This is analogous to having a mechanical valve instantly closing the injectors. When the local boundary pressure falls below the stagnation pressure, the steady quasi-1D nozzle equations are used to calculate the inlet condition. 
Two- and three-dimensional simulations with injectors \cite{schwer2011effect} found that little backflow occurred, supporting the popular model, however, pressure waves penetrating into the injectors and plenum were also observed. Numerical (\textit{e.g.} \cite{schwer2011effect,yanEffectsSlotInjection2021}) and experimental (\textit{e.g} \cite{bedick2019characterization,yang2022suppression,stout2020demonstrated}) investigations of RDE injectors have focused on back flow, mixing, shock transmission into the plenum and engine performance, but the development and impact of shock waves inside these injectors has not been explored in depth. Raman \textit{et al.} \cite{ramanNonidealitiesRotatingDetonation2023} have written a recent review of non-idealities in RDE.

Injector design is confounded by the need to accommodate a continuously changing pressure caused by the rotating detonation. When the injected flow becomes over-expanded, shock waves will appear in the divergent section of the injector and reduce the stagnation pressure. This can have a significant impact on the engine performance. Recent experiments by Baratta and Stout \cite{stout2020demonstrated} with a number of injector types and throat/combustor area ratios all resulted in a net stagnation pressure loss. Shock generation inside the injector was cited as a contributing factor to the pressure loss, but was not directly investigated.

This study introduces a novel way to implement the injectors into numerical simulations. Quasi-two-dimensional area variation is used to implement a generic converging-diverging injector. The simulations are used to assess the importance of the gas dynamic injector--detonation interaction on engine performance. The model is described in section \ref{sec:model}, followed by its numerical implementation in section \ref{sec:numerical method}. The instantaneous and averaged simulation results are presented in section \ref{sec:results}. The flow history along particle paths and stream lines are analyzed and discussed in section \ref{sec:discussion} and conclusions are drawn in section \ref{sec:conclusion}. 

\section{\label{sec:model}Model}

\subsection{Equations and parameters}
The three-dimensional annular combustor is reduced to a rectangular two-dimensional domain \cite{zhdan1990calculation,yi2011propulsive,tsuboi2015numerical,yanEffectsSlotInjection2021} in order to reduce computational costs and simplify the analysis. This is justified when the annulus thickness is much smaller than its diameter, assuming fluctuations in the radial direction are small. 
Comparisons of two- and three-dimensional simulations \cite{tsuboi2015numerical} have shown that they are qualitatively similar.

\begin{figure*}
	\centering
	\begin{subfigure}[t]{0.50\textwidth}
		\centering
		\includegraphics[scale=1]{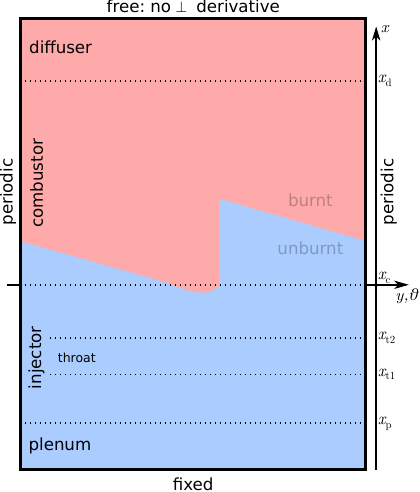}
		\caption{\label{fig:simulation:concept:top}Planar view (fresh/burnt gasses illustrated in blue/red)}
	\end{subfigure}%
	\begin{subfigure}[t]{0.33\textwidth}
		\centering \includegraphics[scale=1]{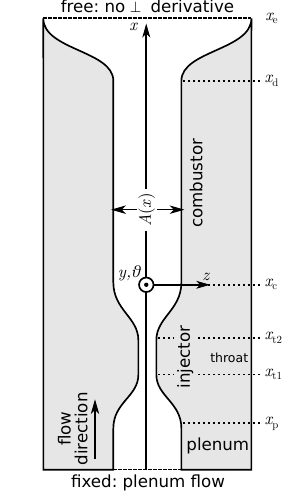}
		\caption{\label{fig:simulation:concept:cross}Quasi-cross-section}
	\end{subfigure}%
	\caption{Illustrations of the quasi-two-dimensional model}
	\label{fig:simulation:concept}
\end{figure*}

The two-dimensional plane is obtained by `unwrapping' the cylindrical annulus. The unwrapped rectangular domain is illustrated in figure \ref{fig:simulation:concept:top}. Reactants flow slowly from the plenum at the bottom into the converging-diverging injector in the axial direction ($x$-axis, vertical).
Reactants are injected into the combustion chamber where the detonation travels azimuthally along the $y$-axis ($\theta$, horizontal). The azimuthal direction is wrapped at the boundaries to mimic the cylindrical shape. Detonation products leave the combustor through a supersonic diffuser at the top.

The injector is modelled as a quasi-two-dimensional converging-diverging nozzle with a long throat. Its transverse cross-section is illustrated in figure \ref{fig:simulation:concept:cross}. The cross-sectional area $A(x)$ is varied axially by changing the channel thickness in the $z$ direction (radial direction in an annular rotating detonation engine). The channel is symmetric about the $x$--$y$ plane and flow is averaged in the $z$ direction. The three-dimensional geometry is simplified to a quasi-two-dimensional geometry modelled with an area (thickness) divergence term $\frac{1}{\hat{A}}\pde{\hat{A}}{\hat{x}}$.

The idea is similar to the classic quasi-one-dimensional nozzle formalism, where the cross-sectional area (radius, thickness) of a nozzle varies along its axis, and so does the flow, which is assumed to be uniform on these cross-sections (in the $r$--$\theta$ plane). 
The quasi-two-dimensional concept differs in that uniformity is only assumed in the radial direction of the annulus, but not in the azimuthal direction. This means the flow field varies in two dimensions, axially ($x$) and azimuthally ($y$). The cross-sectional area (annulus thickness, $z$) variations along the axis are treated in a quasi-one-dimensional manner whilst allowing cross-flow in the azimuthal direction.
No assumptions are made about steadiness nor isentropy. The quasi-two-dimensional concept was previously used to investigate the effect of a divergent outlet nozzle in RDE by Zhdan \textit{et al.} \cite{zhdan2007mathematical,zhdanMathematicalModelContinuous2008}, Fieviosohn \textit{et al.} \cite{fievisohnQuasi2DSimulationsNozzled2018} and Li \textit{et al.} \cite{liNozzleDesignRotating2022} and in a reactive flow calculation by Xiao \textit{et al.} \cite{xiaoEffectBoundaryLayer2021a}. Equations for flow with area variation are covered in section 1.6.2 of Toro's book \cite{toroRiemannSolversNumerical2009}.

The quasi-two-dimensional, compressible, unsteady, reactive Euler equations with area variations in the $x$-direction used to flatten the converging-diverging injector and the diffuser into the two dimensional representation of figure \ref{fig:simulation:concept:top} are
\begin{equation}
\label{eq:governing_equations}
\begin{aligned}
\pde{}{\hat{t}} (\hat{\rho})
+ \pde{}{\hat{x}}(\hat{\rho} \hat{u}) + \pde{}{\hat{y}} (\hat{\rho} \hat{v}) =& - \hat{\rho} \hat{u} \frac{1}{\hat{A}} \pde{\hat{A}}{\hat{x}}
\\
\pde{}{\hat{t}} (\hat{\rho} \hat{u})
+ \pde{}{\hat{x}}(\hat{\rho} \hat{u}^2 + \hat{p}) + \pde{}{\hat{y}} (\hat{\rho} \hat{u} \hat{v}) =& - \hat{\rho} \hat{u}^2 \frac{1}{\hat{A}}\pde{\hat{A}}{\hat{x}}
\\
\pde{}{\hat{t}} (\hat{\rho} \hat{v})
+ \pde{}{\hat{x}}(\hat{\rho} \hat{v} \hat{u}) + \pde{}{\hat{y}} (\hat{\rho} \hat{v}^2 + \hat{p}) =& - \hat{\rho} \hat{v} \hat{u} \frac{1}{\hat{A}}\pde{\hat{A}}{\hat{x}}
\\
\pde{}{\hat{t}} (\hat{\rho} \hat{e}_{\mathrm{tot}}) + \pde{}{\hat{x}}(\hat{\rho} \hat{e}_{\mathrm{tot}} \hat{u} + \hat{p}\hat{u})  + \pde{}{\hat{y}} (\hat{\rho} \hat{e}_{\mathrm{tot}} \hat{v} + \hat{p} \hat{v})
=& \hat{Q} \hat{\rho} \hat{\omega} - (\hat{\rho} \hat{e}_{\mathrm{tot}} + \hat{p}) \hat{u} \frac{1}{\hat{A}} \pde{\hat{A}}{\hat{x}}
\\
\pde{}{\hat{t}} (\hat{\rho} Y)
+ \pde{}{\hat{x}}(\hat{\rho} Y \hat{u}) + \pde{}{\hat{y}} (\hat{\rho} Y \hat{v}) =& \hat{\rho} \hat{\omega} - \hat{\rho} Y \hat{u} \frac{1}{\hat{A}} \pde{\hat{A}}{\hat{x}}
\end{aligned}
\end{equation}
with time $t$, Cartesian coordinates $x$ and $y$, velocity components $u$ and $v$, density $\rho$,  pressure $p$, cross-sectional thickness $A$, heat release $Q$, reaction rate $\omega$, reaction progress variable $Y$ which ranges from 0 in the reactants to 1 in the products, and total energy
\begin{align*}
\hat{e}_{\mathrm{tot}} = \frac{\hat{p}}{\hat{\rho} (\gamma-1)} + \frac{1}{2} (\hat{u}^2 + \hat{v}^2)
\end{align*}
for a calorically perfect gas with isentropic exponent $\gamma$. Circumflex accents represent dimensional values.
A one-step Arrhenius reaction rate
\begin{align*}
\hat{\omega} = \hat{k} (1-Y)\exp \left(- \frac{\hat{E}_{\mathrm{a}}}{\hat{R}\hat{T}}\right)
\end{align*}
is used. The pre-exponential factor $k$ scales the reaction, the activation energy $E_{\mathrm{a}}$ controls the temperature $T$ sensitivity, and $R$ is the gas constant. Temperature and pressure are related through the ideal gas equation of state
\begin{align*}
\hat{p} = \hat{\rho} \hat{R} \hat{T}.
\end{align*}

The variables are non-dimensionalized by a reference state (subscript `ref')
\begin{align*}
\rho = \frac{\hat{\rho}}{\hat{\rho}_{\text{ref}}}, &&
p = \frac{\hat{p}}{\hat{p}_{\text{ref}}}, &&
T = \frac{\hat{T}}{\hat{T}_{\text{ref}}}, && \text{and} &&
x_i = \frac{\hat{x}_i}{\hat{\Delta}_{\frac{1}{2}}}, && 
\end{align*}
chosen to be the injector's supersonic design condition, and the half-reaction length $\hat{\Delta}_{\frac{1}{2}}$ of a Chapman-Jouguet (CJ) detonation propagating through the supersonic flow of the injector.
The other values are non-dimensionalized by
\begin{align*}
\hat{u}_{i,\text{ref}} = \sqrt{\frac{\hat{p}_{\mathrm{ref}}}{\hat{\rho}_{\mathrm{ref}}}},
&& 
\hat{t}_{\text{ref}} = \frac{\hat{x}_{\text{ref}}}{\hat{u}_{\text{ref}}},
&&
\hat{k}_{\mathrm{ref}} = \frac{\hat{u}_{\mathrm{ref}}}{\hat{x}_{\mathrm{ref}}},
&&
\hat{E}_{\mathrm{a,ref}} = \hat{R} \hat{T}_{\mathrm{ref}},
&& \text{and} &&\\
\hat{Q}_{\mathrm{ref}} = \hat{R} \hat{T}_{\mathrm{ref}}
\end{align*}
for the non-dimensional equations to keep the same form as the dimensional equations.

A combustor/injector throat area ratio of ${A_{\mathrm{c}}}/{A_{\mathrm{t}}} = 4$ is used, which falls into the typical range of ${A_{\mathrm{c}}}/{A_{\mathrm{t}}} = 2$ to $10$ used in experiments.
An isentropic exponent of $\gamma = 1.32$, activation energy of $E_{\mathrm{a}} = 50$, and heat release of $Q=56$ are used.
These parameters model a stoichiometric hydrogen and oxygen detonation through the supersonic flow exiting injector with a stagnation temperature of 300 K and 5 atm. 
These parameters were calculated using Cantera \cite{goodwin_cantera_2016} with the Shock and Detonation Toolbox \cite{browne_numerical_2015} and the Stanford H2/O2 mechanism \cite{hongImprovedH2O22011} by isentropically expanding the mixture from stagnation through the injector to the supersonic state. The Chapman-Jouguet detonation Mach number at the injected temperature and pressure was used to calculate heat release using Chapman-Jouguet theory \cite{leeDetonationPhenomenon2008}, $Q = \frac{\gamma}{2(\gamma^2 - 1)} \left(\frac{M_{\mathrm{CJ}}^2 - 1}{M_{\mathrm{CJ}}} \right)^2 $. The activation energy was calculated from the ignition times of constant volume combustion at the von Neumann state by perturbing the initial temperatures. The large values of non-dimensional heat release and activation energy are due to the low temperature of the injected state. The activation energy has been lowered to ensure initiation.

The pre-exponential factor $k$ is chosen so that the non-dimensional half-reaction length is unity, $\Delta_{\frac{1}{2}} = 1$, but as will be described in section \ref{sec:numerical method}, the simulation is not resolved enough to capture the detonation structure. Reactants are converted to products over approximately one grid point, resulting in a Chapman-Jouguet wave.

The Euler equations are commonly used to simulate denotations, which react two or three orders of magnitude faster than deflagrations. Flames and diffusive phenomena, such as those along the contact surface between the injected reactants and detonation products, are not modelled in this study as a consequence; here they are not controlled by physical parameters such as species diffusivity, viscosity or heat conduction. Accurately representing all these length scales is very computationally expensive so state-of-the-art engine-scale simulations typically neglect or only selectively resolve these phenomena.

Without resolving diffusive or reactive scales the model becomes effectively self-similar. This kind of self-similarity is present in other gas dynamics problems such as shock reflection or rocket nozzles, where physical scales exist at various sizes (\textit{e.g.} the viscous shock thickness, equilibrium effects, or heat conduction to walls) but can be ignored at intermediate scales. The length scales presented in the results are arbitrary within the limits of these assumptions.

\subsection{Boundary and initial conditions}

The domain is illustrated in figure \ref{fig:simulation:concept:top}.
Flow enters the domain axially in the plenum at a fixed subsonic state $T_{\mathrm{in}}=2.25251$, $p_{\mathrm{in}}=28.4938$, $u_{\mathrm{in}}=0.254915$. The gas isentropically expands from the plenum through the injector and into the combustor where the detonation propagates azimuthally. Periodic boundary conditions are used in the $y$--direction. Detonation products exit the combustor through a divergent section that terminates at the domain boundary where there are no perpendicular derivatives. The flow is supersonic in the diverging section, ensuring boundary errors do not feed back into the combustor. Reactions are forbidden in the injector and plenum, where fuel and oxidizer are unmixed.

The rectangular domain measures $1200 \Delta_{\frac{1}{2}} \times 600 \Delta_{\frac{1}{2}}$. 
The plenum, injector, combustor, and diffuser lengths are available in table \ref{tab:area variation}.
The cross-sectional areas of the plenum, injector throat, and combustor are constant; the cross-sectional area of the diffuser, convergent and divergent portions of the injector are cosines of $x$, as described in table \ref{tab:area variation}.

\begin{table*}
	\centering
	\caption{\label{tab:area variation}Cross-sectional area variation (see figure \ref{fig:simulation:concept} for an illustration)}
	\begin{tabular}{lrclc}
		\hline \hline
		& \multicolumn{3}{c}{axial position} & cross-sectional area
		\\
		\hline
		exit & \multicolumn{3}{c}{$x = x_{\mathrm{e}} = 1000$} & $A_{\mathrm{e}}=10$
		\\
		diffuser & $950 = x_{\mathrm{d}} $&$ \le x \le $&$ x_{\mathrm{e}} = 1000$ & $a_3 \cos(b_3 (x-c_3) )+d_3$
		\\
		combustor & $0 = x_{\mathrm{c}} $&$ \le x \le $&$ x_{\mathrm{d}} = 950$ & $A_{\mathrm{c}} = 1$
		\\
		injector (diverging) & $-25 = x_{\mathrm{t2}} $&$ \le x \le $&$ x_{\mathrm{c}} = 0$ & $a_2 \cos(b_2 (x-c_2) )+d_2$
		\\
		injector (throat) & $-125 = x_{\mathrm{t1}} $&$ \le x \le $&$ x_{\mathrm{t2}} = -25$ & $A_{\mathrm{t}} = 1/4$
		\\
		injector (converging) & $-150 = x_{\mathrm{p}} $&$ \le x \le $&$ x_{\mathrm{t1}} = -125$ & $a_1 \cos(b_1 (x-c_1) )+d_1$
		\\ 
		plenum & $-200 = x_{\mathrm{inlet}} $&$ \le x \le $&$ x_{\mathrm{p}} = -150$ & $A_{\mathrm{p}} = 1$
		\\
		\hline
		diffuser & \multicolumn{4}{c}{
			\begin{tabular}{cccc}
				$a_3 = \frac{A_{\mathrm{c}} - A_{\mathrm{e}}}{2}$,
				& 
				$b_3 = \frac{\pi }{x_{\mathrm{d}}-x_{\mathrm{e}}}$,
				& 
				$c_3 = x_{\mathrm{d}}$,
				&
				$d_3 =  \frac{A_{\mathrm{c}} + A_{\mathrm{e}}}{2}$
		\end{tabular}} 
		\\
		injector (diverging) & \multicolumn{4}{c}{
			\begin{tabular}{cccc}
				$a_2 = \frac{A_{\mathrm{t}} - A_{\mathrm{c}}}{2}$,
				& 
				$b_2 = \frac{\pi }{x_{\mathrm{t2}}-x_{\mathrm{c}}}$,
				& 
				$c_2 = x_{\mathrm{t2}}$,
				&
				$d_2 =  \frac{A_{\mathrm{c}} + A_{\mathrm{t}}}{2}$
		\end{tabular}} 
		\\
		injector (converging) & \multicolumn{4}{c}{
			\begin{tabular}{cccc}
				$a_1 = \frac{A_{\mathrm{p}} - A_{\mathrm{t}}}{2}$,
				&
				$b_1 = \frac{\pi }{x_{\mathrm{t1}}-x_{\mathrm{p}}}$ ,
				& 
				$c_1 = x_{\mathrm{p}}$,
				&
				$d_1 =  \frac{A_{\mathrm{p}} + A_{\mathrm{t}}}{2}$
		\end{tabular}}
		\\
		\hline \hline
	\end{tabular}
\end{table*}

Initially, values of $p = \rho = 1$ are set in the combustor, $u$ is calculated from the Mach number, and $v=0$. The Mach number is calculated everywhere by solving the area ratio
\begin{align}
\label{eq:steady_quasi-1D-area_ratio}
\frac{A_{\mathrm{t}}}{A} = \frac{\left(\frac{\gamma+1}{2}\right)^{\frac{\gamma+1}{2(\gamma-1)}} M}{\left(1 + \frac{\gamma-1}{2}M^2\right)^{\frac{\gamma+1}{2(\gamma-1)}}}
\end{align}
for Mach number $M$ and choosing the subsonic branch in the plenum and the supersonic branch in the combustor. The state and velocity are calculated assuming isentropic flow. 
The saw-tooth pattern shown in figure \ref{fig:simulation:IC} is initially set with a blast zone within the dashed line at a state calculated using the Rankine-Hugoniot equations for a shock travelling at $1.1$ times the local Chapman-Jouguet velocity. This blast zone was determined through some experimentation to initiate a single detonation wave propagating in the positive $y$ direction.

\begin{Figure}
	\centering
	\includegraphics[scale=1,angle=0]{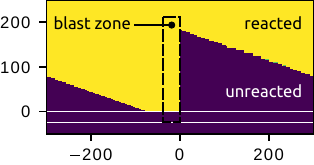}
	\captionof{figure}{Initial conditions of reaction progress and blast zone area (within the dashed line)}
	\label{fig:simulation:IC}
\end{Figure}

\section{\label{sec:numerical method}Numerical Method}

The two-dimensional compressible Euler equations \eqref{eq:governing_equations} are solved using the computational package ``mg'', developed by Falle \cite{falle_self-similar_1991}. An exact Godunov \cite{godunov_difference_1959} scheme with a van Leer flux limiter \cite{van_leer_towards_1977} is used to achieve a second-order solution in space, and a predictor-corrector scheme achieves a second-order solution in time. The chemical reactions and area divergence terms (right-hand side of the governing equation \ref{eq:governing_equations}) are added as sources. Lagrangian point-particles are periodically added at the inlet. Their positions are advected with first-order accuracy.

The equations are discretized over a $48 \times 24$ Cartesian mesh. Adaptive mesh refinement \cite{falle_body_1992} is used to increase accuracy over regions of interest. The resolution is doubled for five grids in each direction where local differences of pressure, density, or velocity exceed $1\%$ between mesh levels, or until four levels of refinement are reached. This provides a maximum grid of $768 \times 384$, or $0.64$ grids per half-reaction lengths in a detonation propagating through the supersonic injector state. A resolution of $\ge 16$ grids per half-reaction length is typically recommended to recover the unsteady cellular structure of detonations. As a result, the detonation structure, detonation instability, the effect of losses, \textit{etc.}, are not addressed in this study. The goal of this investigation is to assess the importance of detonation--injector interaction. Only time-averaged results will be studied. The time step size is chosen using the Courant-Freidrich-Lewy condition ($CFL$) with $CFL \lesssim 0.4$ at each grid level.

\begin{figure*}
	\centering
	\begin{subfigure}[t]{0.33\textwidth}
		\centering
		\includegraphics[scale=1,angle=0]{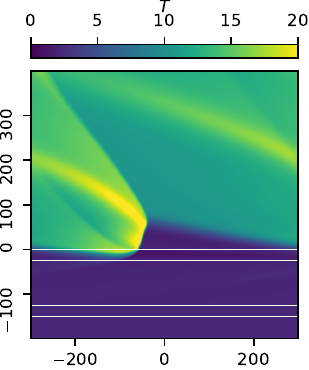}
		\caption{ $0.64 \mathrm{\ grids} / \Delta_{\frac{1}{2}}$}
	\end{subfigure}%
	\begin{subfigure}[t]{0.33\textwidth}
		\centering
		\includegraphics[scale=1,angle=0]{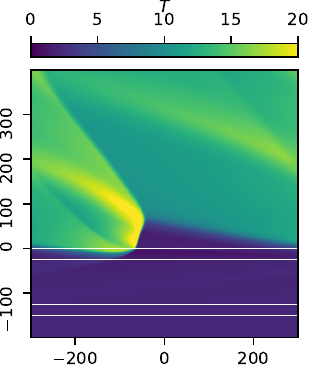}
		\caption{ $1.28 \mathrm{\ grids} / \Delta_{\frac{1}{2}}$}
	\end{subfigure}%
	\begin{subfigure}[t]{0.33\textwidth}
		\centering
		\includegraphics[scale=1,angle=0]{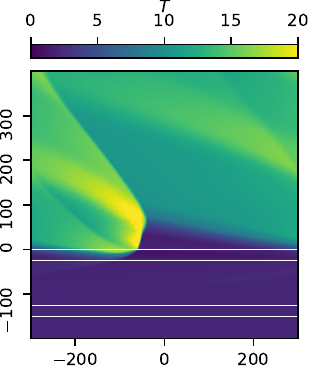}
		\caption{ $2.56 \mathrm{\ grids} / \Delta_{\frac{1}{2}}$}
	\end{subfigure}\\
	\begin{subfigure}[t]{0.66\textwidth}
		\centering
		\includegraphics[scale=1,angle=0]{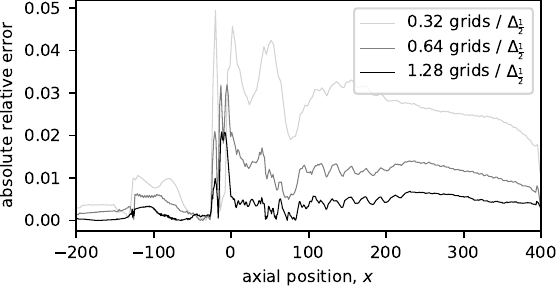}
		\caption{\label{fig:error}Error of stagnation pressure along the axial direction, Favre-averaged around the circumference}
	\end{subfigure}
	\begin{subfigure}[t]{0.33\textwidth}
		\centering
		\includegraphics[scale=1,angle=0]{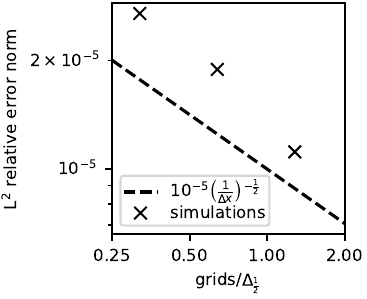}
		\caption{\label{fig:L2norm}Error versus resolution}
	\end{subfigure}
	\caption{The effect of resolution on results: averaged temperature profiles (top) at different resolutions; relative error of the averaged stagnation pressure (bottom)}
	\label{fig:simulation:resolution studies}
\end{figure*}

Figure \ref{fig:simulation:resolution studies} shows the effect of resolution on Favre-averaged results in a shortened domain. The averaged temperature profiles show minor qualitative changes when the resolution is doubled. Differences along the burnt/unburnt interface are attributable to Kelvin-Helmholtz instability. The absolute relative error of Favre-averaged stagnation pressure $p_{\mathrm{stag,lab}}$, a quantity of interest, reduces as resolution is increased. The rate of reduction is shown in figure \ref{fig:L2norm} using the relative error norm $L^2 = \sqrt{\sum(p_{\mathrm{stag,lab}} - p_{\mathrm{stag,lab,ref}})^2/ \sum(p_{\mathrm{stag,lab,ref}})^2}$ over the entire domain, where ``ref'' indicates the most resolved case ($2.56 \mathrm{\ grids} / \Delta_{\frac{1}{2}}$). The error decreases at the expected rate of decay for a second-order-accurate simulation. The portion of the injector blocked by backflow, measured as the distance on the circumference with burnt gas (time averaged, $Y>0.5$) at injector exit ($x=0$), varied non-monotonically with resolution, from 26.2\% to 27.2\% of the circumference.
A maximal resolution of $0.64$ grids per half-reaction lengths is sufficient for this preliminary study. The computation took approximately one hour on 28 Intel Xeon Gold 5120 CPU cores at 2.20 GHz.

Reaction thresholds of $T > 8.333$ and $p > 0.4 p_{\mathrm{vN}}$ need to be met for the reaction to occur, where $p_{\mathrm{vN}}$ is the von Neumann pressure. These values are chosen to lie below the detonation values, as to allow reaction once the flow is shocked, yet sufficiently above the pre-shock state to minimize artificial burning along the product/reactant contact surface due to numerical diffusion, since physical diffusion is not modelled.

Reactions are constrained to the combustor and diffuser. This is analogous to mixing occurring instantaneously once $x \ge 0$, similar to area-averaged boundary conditions which inject pre-mixed reactants directly into the combustor. Mixing could be added to the plenum or injector using complex or simple \cite{whitmanComputationalSimulationMultiheaded2020} models, but is outside the scope of this article.

\section{\label{sec:results}Results}

The azimuthal component of detonation speed measured at the combustor inlet ($x=0$, plotted in figure \ref{fig:simulation:destonation_speed}) has an average of $D_{y\mathrm{,avg}} = 8.85$ over the range $1000 \le t \le 1600$. The detonation propagates at a steady speed once $t \gtrsim 200$, although the downstream flow takes longer to reach its rotating state and there are fluctuations of the detonation speed throughout the simulation. 
The CJ detonation speed through the perfectly expanded injector state is $D_{\mathrm{CJ}} = 9.26$, however, the gas becomes over-expanded in the injector due to elevated combustor pressure.

\begin{Figure}
	\centering
	\includegraphics[scale=1,angle=0]{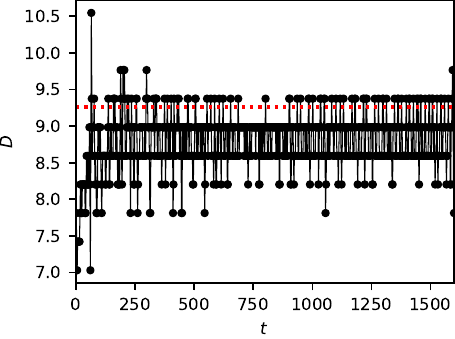}
	\captionof{figure}{Azimuthal component of detonation speed measured at $x=0$ in the lab frame of reference (solid) compared to the CJ detonation speed (dotted)}
	\label{fig:simulation:destonation_speed}
\end{Figure}

\begin{figure*}[ht!]
	\centering
	\begin{subfigure}[t]{0.16\textwidth}
		\centering
		\includegraphics[scale=1,angle=0]{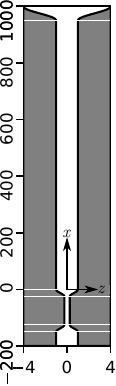}%
		\caption{Cross-section}
	\end{subfigure}%
	\begin{subfigure}[t]{0.28\textwidth}
		\centering
		\includegraphics[scale=1,angle=0]{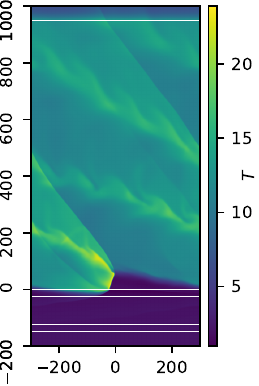}%
		\caption{Tempearture}
	\end{subfigure}%
	\begin{subfigure}[t]{0.28\textwidth}
		\centering
		\includegraphics[scale=1,angle=0]{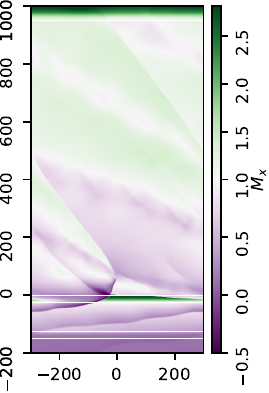}%
		\caption{\label{fig:Mx}Axial Mach number}
	\end{subfigure}%
	\begin{subfigure}[t]{0.29\textwidth}
		\centering
		\includegraphics[scale=1,angle=0]{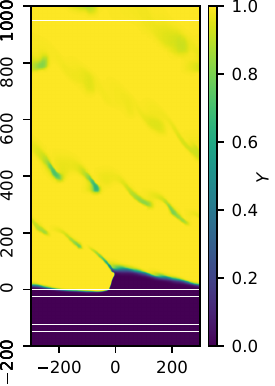}%
		\caption{Reaction progress}
		\label{fig:simulation:instantaneous_results:reaction_progress}
	\end{subfigure}%
	\\
	\begin{subfigure}[t]{0.16\textwidth}
		\centering
		\includegraphics[scale=1,angle=0]{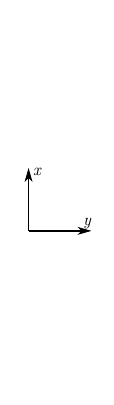}%
	\end{subfigure}%
	\begin{subfigure}[t]{0.28\textwidth}
		\centering
		\includegraphics[scale=1,angle=0]{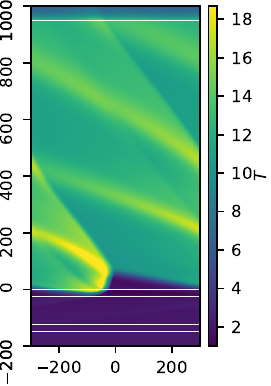}%
		\caption{Tempearture}
		\label{fig:simulation:averaged_results:density}
	\end{subfigure}%
	\begin{subfigure}[t]{0.28\textwidth}
		\centering
		\includegraphics[scale=1,angle=0]{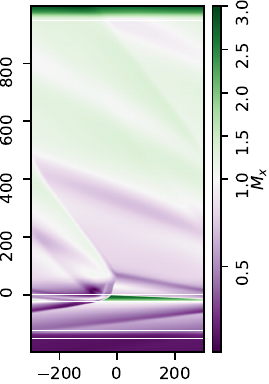}%
		\caption{\label{fig:Mx_avg}Axial Mach number}
	\end{subfigure}%
	\begin{subfigure}[t]{0.28\textwidth}
		\centering
		\includegraphics[scale=1,angle=0]{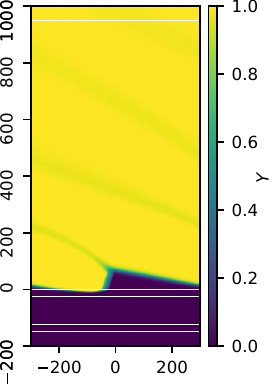}%
		\caption{Reaction progress}
		\label{fig:simulation:averaged_results:reaction_progress}
	\end{subfigure}%
	\caption{Instantaneous (top, $t=1600$) and Favre-averaged results (bottom)}
	\label{fig:simulation:results}
\end{figure*}

Snapshots of the instantaneous flow field at $t=1600$ can be compared to the Favre-time-averaged results in figure \ref{fig:simulation:results}. The averages are obtained by rotating the flow field in the azimuth by $D_{y\mathrm{,avg}} t$, as to line up the detonation fronts, and are then Favre-averaged. Horizontal white lines indicate where area changes begin and end. The instantaneous results have local variations, instabilities along the contact surface separating products from different cycles, and fluctuations in the detonation position that are absent from the averaged result, but the two are otherwise similar.
This means the RDE can be studied in an steady time-averaged manner, in the frame of reference rotating with the detonation. All results presented herein are obtained using Favre-averages.

The axial component of Mach number is plotted in figures \ref{fig:Mx} and \ref{fig:Mx_avg}. Flow exiting the combustor is axially supersonic on average, however, a subsonic patch sonically detached from the detonation exists where the gas is not fully burnt. All flow becomes supersonic at the diverging nozzle.

Figure \ref{fig:simulation:Tavg_zoom} places focus on the flow field near the quasi-two-dimensional injector, the novel aspect of this study. The flow field in the combustor resembles previous numerical studies of RDE in the combustor: a detonation d, travelling from left to right and angled $75^{\circ}$ from the azimuth, and a shock r is reflected from the injector. The detonation transmits a shock t$_2$ into the products of the previous cycle, and a contact layer (between c$_2$ and c$_1$) separates the gasses from different cycles.

\begin{Figure}
	\centering
	\includegraphics[scale=1,angle=0]{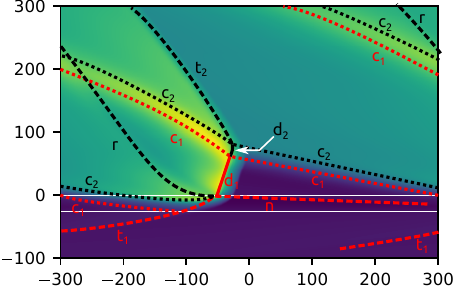}%
	\captionof{figure}{Magnification of the averaged temperature profile near the injector with various features outlined}
	\label{fig:simulation:Tavg_zoom}
\end{Figure}

High pressure behind the detonation forces products into the injector. The injector becomes `unblocked' and reactants begin reentering combustor at $26.6\%$ of the combustor circumference behind the detonation. Much like the shock t$_2$ transmitted into the products, the detonation also transmits a shock t$_1$ through the injector and into the plenum. The thin layer of warm gas formed between contact surfaces c$_1$ and c$_2$ consists of gas processed by the shock t$_1$ in the diverging section of the injector. The warm layer of gas passes through the `upper' portion of the detonation, d$_2$, while the remainder of the reactants are consumed by the `main' portion of the detonation, d$_1$. Reactants from this band periodically form pockets of gas that leave the engine unburnt (figure \ref{fig:simulation:instantaneous_results:reaction_progress}). On average, this appears as a streak of partially reacted gas (figure \ref{fig:simulation:averaged_results:reaction_progress}).

The shock n (nozzle shock) appears in the divergent portion of the injector once t$_1$ enters the injector throat. The shock n forms to match the pressure at the exit of the injector to that of the combustor, which gradually decreases from the detonation pressure to the pre-detonation pressure, but remains above the supersonic design pressure of the injector.

\section{\label{sec:discussion}Discussion}

\begin{figure*}
	\centering
	\begin{subfigure}[t]{0.49\textwidth}
		\centering
		\includegraphics[scale=1,angle=0]{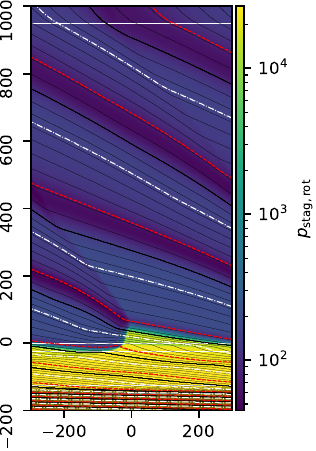}
		\caption{Solid curves: stream lines; thick: $\psi = 0$}
		\label{fig:simulation:streamlines}
	\end{subfigure}\hfill%
	\begin{subfigure}[t]{0.49\textwidth}
		\centering
		\includegraphics[scale=1,angle=0]{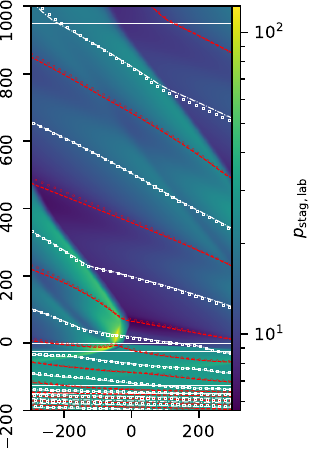}
		\caption{Circles, squares: instantaneous positions of two particles}
		\label{fig:simulation:stagnation_pressure_lab}
	\end{subfigure}%
	\caption{Time-averaged stream lines (left) and two particle paths (right) superimposed onto stagnation pressure in the rotating (left) and lab (right) frames of reference; red dashes: $\psi = 350$; white dot-dashes: $\psi = 1050$}
	\label{fig:simulation:particle_trajectory}
\end{figure*}

The stagnation pressure of a flow, $p_{\mathrm{stag}} = p \left( 1 + \frac{\gamma-1}{2} M^2\right)^{\frac{\gamma}{\gamma-1}}$, is a measure of the flow's ability to do work.
It is dependant on the frame of reference chosen; the stagnation pressure $p_{\mathrm{stag,rot}}$ in the rotating frame of reference, calculated using the Mach number $M = \sqrt{u^2 + (v - D_{y\mathrm{,avg}})^2}/c$, is plotted in figure \ref{fig:simulation:streamlines}, while the stagnation pressure $p_{\mathrm{stag,lab}}$ in the lab frame of reference, calculated using $M = \sqrt{u^2 + v^2}/c$, is shown in figure \ref{fig:simulation:stagnation_pressure_lab}; $c$ is the sound speed.

In both cases, stagnation pressure in the plenum exceeds the stagnation pressure at the exit of the combustor, and is lowest in the flow that passes through the d$_2$ portion of the detonation, where it is not fully reacted. At the exit of the combustor, stagnation pressure calculated using velocities in the lab frame of reference, varies largely depending on its distance from the transmitted shock t$_2$. Additional insight is gained by looking at the stream lines.

A stream-line analysis can be performed due to the development of a steady averaged flow field. The stream lines are found from the contours of the compressible stream function with area variation
\begin{align*}
\psi (x,y) =& \int_{y_0}^{y} (\rho u A) \mathrm{d} y - \int_{x_0}^{x} (\rho (v-D_{y\mathrm{,avg}}) A)_{y_0} \mathrm{d} x
\end{align*}
using the Favre-averaged velocities in the frame of reference rotating with the detonation. The stream lines are plotted as white curves on top of the rotating stagnation pressure in figure \ref{fig:simulation:streamlines}. The stream line $\psi = 0$ is highlighted by the thick black curve. 
Two streamlines will be investigated, one that passes through the upper (d$_2$) portion of the detonation, highlighted with a red dashed line ($- -$, $\psi=350$) in figure \ref{fig:simulation:streamlines}, and another that passes through the main (d$_1$) portion of the detonation, highlighted with a white dot-dashed curve ($\cdot-$, $\psi = 1050$) in figure \ref{fig:simulation:streamlines}. The two stream lines are plotted again in figure \ref{fig:simulation:stagnation_pressure_lab} in black and white, respectively, atop the lab-frame stagnation pressure. Two similar instantaneous particles paths (not time-averaged) are plotted with circles and squares representing the position of their respective Lagrangian particle at one instance in time. The instantaneous particle paths lie close to their corresponding stream lines, accruing some deviation due to small differences between the instantaneous particle history and the time-averaged stream line.

\begin{figure*}
	\centering
	\begin{subfigure}[t]{0.48\textwidth}
		\centering
		\includegraphics[scale=1,angle=0]{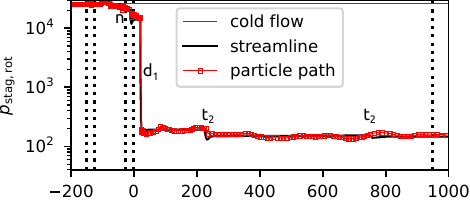}
		\caption{Through the main detonation ($\psi = 1050$); rotating frame}
		\label{fig:simulation:pn=6,sl=1050:stagnation_pressure_rot}
	\end{subfigure}\hfill%
	\begin{subfigure}[t]{0.48\textwidth}
		\centering
		\includegraphics[scale=1,angle=0]{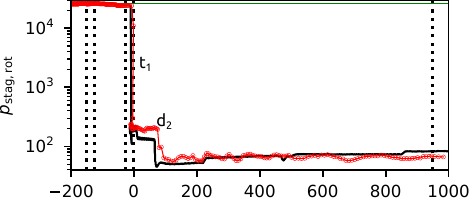}
		\caption{Through the upper detonation ($\psi = 350$); rotating frame}
		\label{fig:simulation:pn=28,sl=350:stagnation_pressure_rot}
	\end{subfigure}
	\\
	\begin{subfigure}[t]{0.48\textwidth}
		\centering
		\includegraphics[scale=1,angle=0]{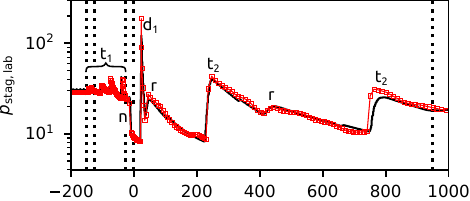}
		\caption{Through the main detonation ($\psi = 1050$); lab frame}
		\label{fig:simulation:pn=6,sl=1050:stagnation_pressure_lab_field}
	\end{subfigure}\hfill%
	\begin{subfigure}[t]{0.48\textwidth}
		\centering
		\includegraphics[scale=1,angle=0]{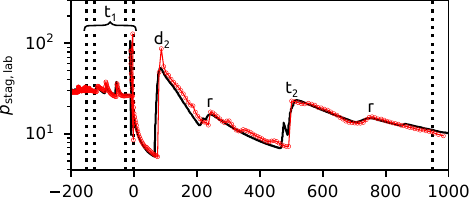}
		\caption{Through the upper detonation ($\psi = 350$); lab frame}
		\label{fig:simulation:pn=28,sl=350:stagnation_pressure_lab}
	\end{subfigure}
	\caption{Stagnation pressure of particles and time-averaged stream lines, plotted against axial position; dotted lines indicated axial locations of area change}
	\label{fig:simulation:streamline_and_particle_stagnation_pressures}
\end{figure*}

The stagnation pressures measured along these stream lines and particle paths are plotted in figure \ref{fig:simulation:streamline_and_particle_stagnation_pressures} against their axial position in the engine. First consider stream line $\psi = 1050$ in figure \ref{fig:simulation:pn=6,sl=1050:stagnation_pressure_rot} which passes through the main portion of the detonation.
The stagnation pressure in the rotating frame of reference of the stream line is plotted in black and the points show the instantaneous value experienced by the Lagrangian particle. There is good agreement between the two. The thin green line represents the stagnation pressure in the inert case (`cold flow', also the initial conditions outside of the initial blast zone), and is constant because the non-reactive case is isentropic. As expected from a steady stream line analysis, the stagnation pressure drops every time the stream line crosses a shock or the detonation. The stream line passing through the upper portion of the detonation, figure \ref{fig:simulation:pn=28,sl=350:stagnation_pressure_rot}, experiences a large stagnation pressure drop in the injector, due to the transmitted shock t$_1$, and a second large drop across the detonation d$_2$.

The rotating component of stagnation pressure, however, does not contribute to the engine's ability to do work. 
Looking instead at the lab-frame stagnation pressure along the stream line, in figure \ref{fig:simulation:pn=6,sl=1050:stagnation_pressure_lab_field},
the stagnation pressure in the plenum is elevated slightly above the initial conditions due to the transmitted shock t$_1$. The stagnation pressure jumps each time the stream line crosses the t$_1$ shock, followed by a decay to near its pre-shock pressure throughout the plenum and injector. The rise and fall of lab-frame stagnation pressure along the stream line is due to the obliqueness and unsteadiness of the shocks and expansion waves to the flow in this frame of reference. The jumps and decays become larger as the stream line approaches the combustor ($x=0$). As the flow accelerates through the divergent portion of the injector, the stream line is shocked by the nozzle shock n, to match the combustor pressure. Stagnation pressure plummets across this shock.
The stream line enters the combustor and sees a large jump in stagnation pressure across the main detonation, which is unsteady in this frame of reference, followed by a gradual decrease due to the unsteady expansion behind. This is repeated each time the stream line crosses an oblique shock, t$_2$ or r. The stream line exits the combustor with a greater stagnation pressure than with which it entered the combustor, a virtue of pressure-gain combustion. However, the stagnation pressure at the exit is lower than the flow entering the plenum, resulting in a net stagnation pressure loss.
The shock in the diverging section of the injector is the culprit, meaning proper injector design is of paramount importance if net pressure-gain combustion is to be achieved.

The lab-frame stagnation pressure history of stream line $\psi = 350$ passing through the upper portion of the detonation ({d}$_{\mathrm{2}}$) is presented in figure \ref{fig:simulation:pn=28,sl=350:stagnation_pressure_lab}. Its history through the plenum, injector throat, and combustor is similar to the previous case. However in the injector's diffuser, the particle is shocked by the transmitted shock {t}$_1$, instead of the nozzle shock {n}$_1$, momentarily forcing it back into the injector. The stream line ends up in the warm strip of gas between {c}$_1$ and {c}$_2$ and expands to a lower pressure than the previous case before crossing the detonation. The stagnation pressure amplification across the detonation is weaker than in the previous case. The stream line undergoes jumps in stagnation pressure across the detonation and oblique shocks, followed by gradual decreases in stagnation pressure, and exits the combustor with half the stagnation pressure of the previous case. A large portion of the difference is due to its relative position to the transmitted shock, t$_2$. If the combustor ended at $x=500$, for example, stream line $\psi =350$ would exit the combustor closer behind the oblique shocked t$_2$ and at a more favourable stagnation pressure. However, figure \ref{fig:simulation:stagnation_pressure_lab} clearly shows that this stream line has a lower stagnation pressure on average.

The two stream lines/particle paths presented are representative of two families of stream tubes, one that passes through the main detonation and the other through the upper detonation.
Stream lines which cross the main detonation consist of roughly $75\%$ of the stream lines in the engine and all follow a thermodynamic cycle (\textit{i.e.} $p$--$v$ diagram) close to the idealized detonation cycle (Fickett-Jacobs).
The stream lines that cross the upper detonation follow a similar thermodynamic cycle, but have a wider variation of states.

\section{\label{sec:conclusion}Conclusion}

A new method of implementing the inlet of a rotating detonation engine was presented. This method uses a quasi-two-dimensional approach to model a generic converging-diverging injector. It is well posed, easily implemented, and can also be used to model other area changes in the engine, such as the diffuser nozzle.

A two-dimensional numerical simulation of a rotating detonation combustor and injector was performed. A periodic solution to the flow in the rotating detonation engine was achieved, with minor perturbations, allowing a steady analysis to be performed using Favre-averaged results in a moving frame of reference. Two families of stream lines were identified. The larger family, consisting of $\sim 75\%$ of stream lines, cross a shock wave in the divergent portion of the injector that causes an irrecoverable loss in lab-frame stagnation pressure. The stagnation pressure gain in the combustor is insufficient to make up for the loss and the flow leaves the combustor with a stagnation pressure lower than in the plenum. The second family of stream lines passes through the upper portion of the detonation and leaves the combustor partially reacted with even less stagnation pressure.

A shock wave appears inside the diverging portion of the injector to compress the overexpanded injector flow to the elevated combustor pressure. This shock wave is an important contributor to stagnation pressure loss. Ignoring the gas dynamic losses in the injector can lead to erroneous conclusions about engine performance. The effect of throat/combustor area ratio will be studied in follow-on work using the quasi-two-dimensional approach presented here.

The case studied in this paper corresponds to a choked, supersonic injector, however, the quasi-two-dimensional approach can also be used to investigate subsonic and unchoked injectors by varying the plenum and exit conditions or area ratios.

\section*{Acknowledgments}

The authors thank Sam Falle from the University of Leeds for generously allowing the use of his code, mg, with which the simulations were performed.

The work was supported by the Aerospace Research Centre of National Research Council of Canada (NRC) under the Aerospace Future Initiative program (AFI) and the Office of Energy Research and Development (OERD) of Natural Resources Canada.

\bibliography{references}

@inproceedings{schwer2011effect,
	title = {Effect of Inlet on Fill Region and Performance of Rotating Detonation Engines},
	booktitle = {AIAA Paper 2011-6044},
	author = {Schwer, Douglas and Kailasanath, K},
	year = {August 2011},
	doi = {https://doi.org/10.2514/6.2011-6044}
}

@inproceedings{schwer2012feedback,
	title = {Feedback into Mixture Plenums in Rotating Detonation Engines},
	author = {Schwer, Douglas and Kailasanath, Kailas},
	booktitle = {AIAA Paper 2012-0617},
	year ={January 2012},
	doi = {https://doi.org/10.2514/6.2012-617}
}

@article{yi2011propulsive,
	title = {Propulsive Performance of a Continuously Rotating Detonation Engine},
	volume = {27},
	number = {1},
	journal = {Journal of Propulsion and Power},
	author = {Yi, Tae-Hyeong and Lou, Jing and Turangan, Cary and Choi, Jeong-Yeol and Wolanski, Piotr},
	year = {2011},
	pages = {171-181},
	doi = {https://doi.org/10.2514/1.46686}
}

@article{tsuboi2015numerical,
	title = {Numerical Estimation of the Thrust Performance on a Rotating Detonation Engine for a Hydrogen--Oxygen Mixture},
	volume = {35},
	number = {2},
	journal = {Proceedings of the Combustion Institute},
	author = {Tsuboi, Nobuyuki and Watanabe, Yusuke and Kojima, Takayuki and Hayashi, A Koichi},
	year = {2015},
	pages = {2005-2013},
	doi = {https://doi.org/10.1016/j.proci.2014.09.010}
}

@article{zhdan2007mathematical,
  title = {Mathematical Modeling of a Rotating Detonation Wave in a Hydrogen-Oxygen Mixture},
  volume = {43},
  number = {4},
  journal = {Combustion, explosion, and shock waves},
  author = {Zhdan, Sergey A and Bykovskii, Fedor A and Vedernikov, Evgenii F},
  year = {2007},
  pages = {449-459},
  doi = {https://doi.org/10.1007/s10573-007-0061-y}
}

@book{goodwin_cantera_2016,
	title = {Cantera: {{An Object}}-Oriented {{Software Toolkit}} for {{Chemical Kinetics}}, {{Thermodynamics}}, and {{Transport Processes}}},
	author = {Goodwin, David G. and Moffat, Harry K. and Speth, Raymond L.},
	year = {2016},
	note = {Version 2.2.1}
}

@techreport{browne_numerical_2015,
	type = {{{GALCIT Report}}},
	title = {Numerical {{Solution Methods}} for {{Shock}} and {{Detonation Jump Conditions}}},
	number = {FM2006.006},
	institution = {{California Institute of Technology}},
	author = {Browne, S. and Ziegler, J. and Shepherd, J.E.},
	year = {2015}
}

@article{falle_self-similar_1991,
	title = {Self-Similar Jets},
	volume = {250},
	number = {3},
	journal = {Monthly Notices of the Royal Astronomical Society},
	author = {Falle, SAEG},
	year = {1991},
	pages = {581--596},
	doi = {https://doi.org/10.1093/mnras/250.3.581}
}

@article{falle_body_1992,
	title = {Body Capturing Using Adaptive {{Cartesian}} Grids},
	journal = {Numerical methods for fluid dynamics},
	author = {Falle, SAEG and Giddings, JR},
	year = {1992},
	pages = {335--342}
}

@article{godunov_difference_1959,
	title = {A Difference Method for Numerical Calculation of Discontinuous Solutions of the Equations of Hydrodynamics},
	volume = {89},
	number = {3},
	journal = {Matematicheskii Sbornik},
	author = {Godunov, Sergei Konstantinovich},
	year = {1959},
	pages = {271--306}
}

@article{van_leer_towards_1977,
	title = {Towards the Ultimate Conservative Difference Scheme. {{IV}}. {{A}} New Approach to Numerical Convection},
	volume = {23},
	number = {3},
	journal = {Journal of computational physics},
	author = {Van Leer, Bram},
	year = {1977},
	pages = {276--299},
	doi = {https://doi.org/10.1006/jcph.1997.5704}
}

@article{heiser2002thermodynamic,
	title = {Thermodynamic Cycle Analysis of Pulse Detonation Engines},
	volume = {18},
	number = {1},
	journal = {Journal of Propulsion and Power},
	author = {Heiser, William H and Pratt, David T},
	year = {2002},
	pages = {68-76},
	doi = {https://doi.org/10.2514/2.5899}
}

@article{zhdan1990calculation,
	title = {Calculation of the Flow of Spin Detonation in an Annular Chamber},
	volume = {26},
	number = {2},
	journal = {Combustion, Explosion, and Shock Waves},
	author = {Zhdan, Sergey A and Mardashev, AM and Mitrofanov, VV},
	year = {1990},
	pages = {210-214},
	doi = {https://doi.org/10.1007/BF00742414}
}

@inproceedings{stout2020demonstrated,
  title = {Demonstrated Low Loss and Low Equivalence Ratio Operation of a Rotating Detonation Engine for Power Generation},
  booktitle = {AIAA Paper 2020-1173},
  author = {Stout, Jeffrey B and Baratta, Alexander},
  year = {January 2020},
  doi = {https://doi.org/10.2514/6.2020-1173}
}

@article{yanEffectsSlotInjection2021,
  title = {Effects of Slot Injection on Detonation Wavelet Characteristics in a Rotating Detonation Engine},
  author = {Yan, Chian and Teng, Honghui and Ng, Hoi Dick},
  year = {2021},
  journal = {Acta Astronautica},
  volume = {182},
  pages = {274--285},
  doi = {https://doi.org/10.1016/j.actaastro.2021.02.010}
}

@article{bedick2019characterization,
  title = {Characterization of Rotating Detonation Engine Injector Response Using Laser-Induced Fluorescence},
  author = {Bedick, Clinton and Ferguson, Don and Strakey, Peter},
  year = {2019},
  journal = {Journal of Propulsion and Power},
  volume = {35},
  number = {4},
  pages = {827--838},
  publisher = {{American Institute of Aeronautics and Astronautics}},
  doi={https://doi.org/10.2514/1.B37309}
}

@article{yang2022suppression,
  title={Suppression of pressure feedback of the rotating detonation combustor by a Tesla inlet configuration},
  author={Yang, Xingkui and Song, Feilong and Wu, Yun and Guo, Shanguang and Xu, Shida and Zhou, Jianping and Liu, Hao},
  journal={Applied Thermal Engineering},
  volume={216},
  pages={119123},
  year={2022},
  doi={https://doi.org/10.1016/j.applthermaleng.2022.119123}
}

@article{ramanNonidealitiesRotatingDetonation2023,
  title = {Nonidealities in Rotating Detonation Engines},
  author = {Raman, Venkat and Prakash, Supraj and Gamba, Mirko},
  year = {2023},
  journal = {Annual Review of Fluid Mechanics},
  volume = {55},
  doi = {https://doi.org/10.1146/annurev-fluid-120720-032612}
}

@article{xiaoEffectBoundaryLayer2021a,
  title = {Effect of {{Boundary Layer Losses}} on {{2D Detonation Cellular Structures}}},
  author = {Xiao, Qiang and Sow, Aliou and Maxwell, Brian McN and Radulescu, Matei Ioan},
  year = {2021},
  journal = {Proceedings of the Combustion Institute},
  volume = {38},
  number = {3},
  pages = {3641--3649},
  doi = {https://doi.org/10.1016/j.proci.2020.07.068}
}

@book{toroRiemannSolversNumerical2009,
  title = {Riemann Solvers and Numerical Methods for Fluid Dynamics: A Practical Introduction},
  author = {Toro, Eleuterio F.},
  year = {2009},
  edition = {Third},
  publisher = {{Springer Berlin Heidelberg}},
  address = {{New York}},
  doi = {https://doi.org/10.1007/b79761}
}

@inproceedings{fievisohnQuasi2DSimulationsNozzled2018,
	title = {Effect of Inlet on Fill Region and Performance of Rotating Detonation Engines},
	booktitle = {AIAA Paper 2018-0881},
	author = {Fievisohn, R and Hoke, J L and Schauer, F R},
	year = {January 2018},
	doi = {https://doi.org/10.2514/6.2018-0881},
}

@article{liNozzleDesignRotating2022,
  title = {Nozzle {{Design}} for {{Rotating Detonation Engine}}},
  author = {Li, Rui and Xu, Jinglei and Huang, Shuai},
  year = {2022},
  journal = {Journal of Propulsion and Power},
  volume = {38},
  number = {5},
  pages = {849--865},
  doi = {https://doi.org/10.2514/1.B38539},
}

@article{zhdanMathematicalModelContinuous2008,
  title = {Mathematical Model of Continuous Detonation in an Annular Combustor with a Supersonic Flow Velocity},
  author = {Zhdan, Sergey A},
  year = {2008},
  journal = {Combustion, Explosion, and Shock Waves},
  volume = {44},
  number = {6},
  pages = {690--697},
  doi = {https://doi.org/10.1007/s10573-008-0104-z},
}

@article{tangThreedimensionalNumericalInvestigations2015,
  title = {Three-Dimensional Numerical Investigations of the Rotating Detonation Engine with a Hollow Combustor},
  author = {Tang, Xin-Meng and Wang, Jian-Ping and Shao, Ye-Tao},
  year = {2015},
  journal = {Combustion and Flame},
  volume = {162},
  number = {4},
  pages = {997--1008},
  doi = {https://doi.org/10.1016/j.combustflame.2014.09.023}
}

@article{nakagamiExperimentalVisualizationStructure2017,
  title = {Experimental Visualization of the Structure of Rotating Detonation Waves in a Disk-Shaped Combustor},
  author = {Nakagami, Soma and Matsuoka, Ken and Kasahara, Jiro and Kumazawa, Yoshiki and Fujii, Jumpei and Matsuo, Akiko and Funaki, Ikkoh},
  year = {2017},
  journal = {Journal of Propulsion and Power},
  volume = {33},
  number = {1},
  pages = {80--88},
  doi = {https://doi.org/10.2514/1.B36084},
}

@article{yamaguchiSupersonicCombustionInduced2019a,
  title = {Supersonic Combustion Induced by Reflective Shuttling Shock Wave in Fan-Shaped Two-Dimensional Combustor},
  author = {Yamaguchi, Masato and Matsuoka, Ken and Kawasaki, Akira and Kasahara, Jiro and Watanabe, Hiroaki and Matsuo, Akiko},
  year = {2019},
  journal = {Proceedings of the Combustion Institute},
  volume = {37},
  number = {3},
  pages = {3741--3747},
  doi = {https://doi.org/10.1016/j.proci.2018.06.210},
}

@article{hongImprovedH2O22011,
  title = {An Improved {{H2}}/{{O2}} Mechanism Based on Recent Shock Tube/Laser Absorption Measurements},
  author = {Hong, Zekai and Davidson, David F and Hanson, Ronald K},
  year = {2011},
  journal = {Combustion and Flame},
  volume = {158},
  number = {4},
  pages = {633--644},
  doi = {https://doi.org/10.1016/j.combustflame.2010.10.002}
}

@book{leeDetonationPhenomenon2008,
  title = {The Detonation Phenomenon},
  author = {Lee, John H. S.},
  year = {2008},
  publisher = {{Cambridge University Press}},
  address = {{Cambridge}},
  doi = {https://doi.org/10.1017/CBO9780511754708}
}

@inproceedings{whitmanComputationalSimulationMultiheaded2020,
  title = {Computational Simulation of Multiheaded Detonation Dynamics in Rotating Detonation Engines},
  booktitle = {AIAA 2020-3877},
  author = {Whitman, Charles R and Ajaero, Chibuikem U and {Connolly-Boutin}, Sean and L'esp{\'e}rance, Xavier and Kiyanda, Charles and Higgins, Andrew J},
  year = {August 2020},
  doi = {https://doi.org/10.2514/6.2020-3877}
}
\end{multicols}
\end{document}